\documentstyle[aps,pre,preprint]{revtex}
\newcommand{\bm}[1]{{\mbox{\boldmath$#1$}}}
\begin{document}

\bibliographystyle{unsrt}

\title{
Nuclear magnetic susceptibility of metals with magnetic impurities 
}
\author{A. M. Dyugaev$^{1,2}$, Yu. N. Ovchinnikov$^{1,2}$, P. Fulde$^1$ and K. Kladko$^1$}
\address{
$^1$ Max-Planck-Institute for Physics of Complex Systems, \\ 
Bayreuther Str. 40 Haus 16,  D-01187 Dresden, Germany \\
$^2$ L. D. Landau Institute for Theoretical Physics, \\
Academy of Sciences of Russia,
Kosygin Str. 2, Moscow, 117940 Russia
}

\maketitle
\begin{abstract}
We consider the contribution of magnetic impurities to the nuclear magnetic susceptibility $\chi$ and 
to the specific heat $C$ of a metal. The impurity contribution to the magnetic susceptibility
has  a $1/T^2$ behaviour, and the impurity contribution to the specific heat has a $1/T$ behaviour,
both in an extended  region of temperatures $T$. In the case of a dirty metal the RKKY interaction  of nuclear spins
and impurity spins is
suppressed for low temperatures and the main contribution to $C$ and $\chi$ is given by their dipole-dipole
interaction. 
\end{abstract}

PACS numbers: 72.15.Eb 

The magnetic impurity contribution to the nuclear specific heat of a solid was established in the paper \cite{dyugaev97}.
In that paper the magnetic dipole-dipole interaction between the nuclear magnetic moments $\mu_n$ and the magnetic moments
of impurities 
$\mu_{imp}$ was considered. This interaction has the form
 \begin{equation}
\label{1}
V_{d.-d.} = \frac {{\bm{\mu}}{_1}{ \bm {\mu}}_2  r^2 - 3 ({\bm {\mu}}{_1} {\bf r} ) ( {\bm {\mu}}{_2} {\bf r})}{r^5}.
\end{equation}

The interaction (\ref{1}) takes place both in metals and insulators. Experiments
\cite{wendler}, \cite{herrmann1},\cite{herrmann2}
study the metal case.
 For metals the RKKY interaction is present in addition to (\ref{1}), 

\begin{equation}
\label{2}
V_{RKKY} = - \kappa \frac{ {\bm{\mu}}_n {\bm{\mu}_{imp}}}{r^3} \cos (2 k_F r ) 
 e^{- r / r_0}, 
\mbox { for  $k_F r \gg 1$}.
\end{equation}

The characteristic length $r_0$ describes a suppression of the RKKY interaction by nonmagnetic impurities. It is of the order of the electron mean free path and is inversely proportional to the concentration of the nonmagnetic impurities. The value of the parameter
$\kappa$ defines the relative weight of the RKKY interaction (\ref{2}) related to the universal interaction (\ref{1}), that 
does not depend on the impurity concentration. The parameter $\kappa$  can  not be found precisely. It is known only,
that $\kappa$ increases monotonously with an increase of the nuclear charge $Z$, leading to  $\kappa \gg 1$ for $Z \gg 1$. It was shown in the paper \cite{dyugaev97}, that at low temperatures the main part of the magnetic impurity contribution to the  nuclear specific heat comes from  nuclei situated at large distances from the  magnetic impurity. Increasing the concentration of nonmagnetic impurities one can suppress the RKKY interaction and come to the situation, where  only the interaction (\ref{1}) is important. Then an interesting possibility appears to derive the value of the parameter $\kappa$ in (\ref{2}) by comparing the 
 values of $C$ and $\chi$ for clean and dirty metals. 

For the sake of simplicity we consider the case, where the value of the nuclear spin is $S_n = 1/2$, which  is valid for the
compound $PtFe_x$ experimentally investigated in \cite{wendler}. $^{195}Pt$ is the only stable isotope of platinum that has a nonzero spin,  $S=1/2$, $\mu = 0.6095 \mu_{N} $. The naturally available  $Pt$ is an isotopic  mixture, containing about 34\% of  $^{195}Pt$. According 
to \cite{wendler} the temperature of the nuclear spin spin-glass  transition is definitely lower than $10^{-6}K$.   
The temperature of the impurity spin spin-glass  transition is much higher, lying in the interval $T_{Fe}= (2.7-8.5)10^{-3} K$ for $Fe$ impurities.
There exists hence an extended region of temperatures $T_{cn} << T << T_{Fe}$.  In this region  the impurity ($Fe$) spins are frozen and do not
contribute to $C$ and $\chi$, contrary to the nuclear spins, which are uncorrelated. In the first approximation for the aforementioned temperature
region we can put $T_{cn}=0$ and $T_{Fe}= \infty$.  Then the two systems, the nuclear spin system and the impurity spin system, considered
separately, do not posses characteristic temperature scales. A characteristic temperature scale $\tau$ appears only after considering the interactions
(\ref{1}, \ref{2})
between  the two systems. 

To find the energy scale $\tau$,  let us first consider an effective magnetic field ${\bf H}_n$ acting on a nuclear spin. This field is a sum of an external
magnetic field ${\bf H}_0$ and an internal magnetic field ${\bf H}_1({\bf r}_n)$.

\begin{equation}
\label{3}
{ \bf H}_n = {\bf H}_0 + {\bf H}_1({\bf r}_n)
\end{equation}

The free energy $f_n$ of a single nucleus with a $S=1/2$ spin in the magnetic field $H_n$ is
\begin{equation}
f_n = -T \ln [ 2 \cosh  \frac{\mu_n H_n}{2T}].
\end{equation}

The magnetic moment of the nucleus ${\bf m}_n$ and the contribution of the nucleus to the heat capacitance
$C_n$ are then given by a differentiation of $f_n$
\begin{eqnarray}
\label{4}
{\bf m}_n = - {\frac{ \partial  f_n}{\partial {\bf H}_n }} = 
\frac{\mu_n}{2} \frac{{\bf H}_n}{H_n}  \tanh \frac {\mu_n H_n }{2 T}, \\ 
\nonumber
C_n = - \beta^2 \frac{\partial^2 f_n}
{\partial \beta^2}  = 
(\frac {\mu_n H_n }{2 T})^2 \frac {1}{\cosh^2 \frac{ \mu_n H_n }{2 T}},  \; \beta =1/T. 
\end{eqnarray}

Performing a sum of $C_n$ and ${\bf m}_n$ over all nuclei one finds the magnetisation moment ${\bf M}$ and the 
heat capacity $C$ per unit volume. 
It is then convenient to separate out the contribution of a clean metal.

\begin{eqnarray}
\label{5}
{\bf M} = {\bf M}_0 + \frac{1}{V} \sum_n ({\bf m}_n - {\bf m}_0), \\ \nonumber
C = C_0 + \frac {1}{V} \sum_n (C_n -C_{n0})
\end{eqnarray}

The values of ${\bf m}_0$ and $C_{n0}$ are defined by the formula  (\ref{4}) taking  ${\bf H}_n = {\bf H}_0$.

For  a low concentration of magnetic impurities their influence regions do not overlap, and one can consider a single
impurity placed at the origin of coordinates. A contribution to $C$ and ${\bf \chi}$ comes from a region near the 
magnetic impurity, which is  much larger than the lattice spacing, so the summations can be replaced by  integrations with respect to
$d{\bf r}$. The magnetic  impurity spins are in a spin-glass state and  have  random directions with respect to the external magnetic
field, therefore  one needs to average the resulting expressions with respect to the direction of the magnetic impurity spin.

Let us consider the absolute values of ${\bf M}$ and ${\bf M}_0$

\begin{equation}
\label{6}
{\bf M} = M \frac{{\bf H}_0}{H_0}, \; 
 {\bf M}_0 = M_0 \frac{{\bf H}_0}{H_0}. 
\end{equation}

Then 
\begin{eqnarray}
\label{7}
M = M_0 + \frac {\mu_n}{4} n_{imp} n_n,
 \int^1_{-1} dx \int d {\bf r} 
 [\frac { {\bf H} {\bf H}_0}{H_n H_0} \tanh \frac {\mu_n H_n }{2 T}
 - \tanh \frac {\mu_n H_0}{ 2 T} ], \\
\nonumber
C = C_0 + \frac{1}{2} n_n n_{imp} \int^1_{-1} dx \int {\bf dr} (\frac{\mu_n H_n}
{2 T})^2 \frac{1}{\cosh^2 
\frac {\mu_n H_n}{ 2 T}}
-
 (\frac{\mu_n H_0}
{2 T})^2 \frac{1}{\cosh^2
\frac {\mu_n H_0}{ 2 T}}
.
\end{eqnarray}
Here $x = \cos \theta $, $\theta$ is the angle between ${\bf H}_0$ and ${\bf H}_1$ in (\ref{3}).
Differentiating $M$ with respect to $H_0$ (\ref{6}) and taking  $H_0 = 0$, one finds the impurity
contribution to the nuclear magnetic susceptibility at zero magnetic field, a  characteristic
usually measured experimentally.

\begin{eqnarray}
\label{8}
\chi = \chi_0 [ 1 - J(T) ], \;  \chi_0 = \frac{\mu_n^2 n_n }{4T}, \\ \nonumber
J(T) = n_{imp} \int d{\bf r} [ \frac{1}{3} \tanh^2 \frac {\mu_n H_1}{2 T} + \frac{2}{3} (1- 
\frac{2 T}{ \mu_n H_1} \tanh \frac{\mu_n H_1}{2T}) ] 
\end{eqnarray}

Let us first consider the case where the 
magnetic dipole-dipole interaction (\ref{1}) plays the major role and the
RKKY interaction can be neglected. 
In this case the expressions for the temperature dependence 
of the specific heat $C$ were derived in (\ref{1}). For the magnetic susceptibility $\chi$  
an  integration with respect to $d{\bf r}$ in (\ref{8}) gives after some mathematics 
the value of $J$, which we denote as $J_1$
\begin{eqnarray}
\label{9}
J_1(T) = \frac{\theta_1}{T}, \\ \nonumber
\theta_1 = \frac {2 \pi} {3} \mu_{imp} \mu_n n_{imp} [ 1 + \frac{1}{2 {\sqrt 3}}
\ln (2 + {\sqrt 3}) ] I_1, \\ \nonumber
I_1 = \int^{\infty}_0 \frac{dz}{z^2} [ \frac{1}{3} \tanh^2 z + \frac{2}{3} (1 - \frac{\tanh z}{z}) ] = 1.137.
\end{eqnarray}

As a result the impurity contribution to $\chi$ is proportional to $1/T^2$ (\ref{8}, \ref{9}). The magnetic susceptibility of
a clean metal  has one more $1/T^2$  correction,
 related to the  spin-spin interaction among nuclei. The 
expansion of $\chi$ in powers of $1/T$ has then  the form

\begin{equation}
\label{10}
\chi = \frac {\mu_n^2 n_n}{4 T} ( 1- \frac{\theta}{T}),  \; \theta = \theta_n +\theta_1 .
\end{equation}

One can roughly estimate the value of $\theta_n$  as $\theta_n \sim \mu^2_n n_n$.  A relation of  $\theta_1$ to
$\theta_n$  contains a small parameter
$x_{imp} = \frac{n_{imp}}{n_n}$, but also a large parameter $\frac{\mu_{imp}}{\mu_n} \sim 10^4$. Therefore, already for a small
concentration of magnetic impurities $x_{imp} \sim 10^{-4}$, the impurity
part  makes the main contribution to $\theta$ in (\ref{10}).

Let us consider now another limiting case, where the RKKY interaction is much
stronger than the magnetic dipole interaction. 
For the RKKY interaction  (\ref{2}) using  the expression for the impurity field $H_1(r)$
\begin{equation}
\label{11}
H_1(r)  = \kappa \frac{ \mu_{imp}}{r^3} \cos (2 k_F r )  e^{- r / r_0}, 
\end{equation}
and performing the  substitution  of the integration variable in (\ref{7},\ref{8}) $r \rightarrow z $, where  
\begin{equation}
z = \frac{\mu_{imp}\mu_n \kappa}{ 2 T r^3},
\end{equation}
one gets
\begin{eqnarray}
\label{12}
\chi(T) = \chi_0 [ 1 - J_2 (T) ], \\ \nonumber
J_2(T) = \frac {\theta_2 (T)}{T},  \; \;  \theta_2 (T) = \frac{2 \pi}{3} \mu_{imp} \mu_n
\kappa n_{imp} I_2(T), \\ \nonumber
I_2 (T) =  \int^{\infty}_0 \frac{dz}{z^2} [ \frac{1}{3} \tanh^2 y + \frac{2}{3} (1 - \frac{\tanh y}{y}) ],
\end{eqnarray}
where 
\begin{equation}
\label{13}
y = y(z) = z \cos (\frac{z_0}{z})^{1/3} \exp [ - (\frac{z_1}{z})^{1/3} ].
\end{equation}
The parameters $z_0, z_1$ are related to the  two characteristic temperature scales
\begin{eqnarray}
\label{14}
z_0 = \frac{T_0}{T}, \;\; T_0 = 12 \pi^2 n_e \mu_n \mu_{imp} \kappa ,\\ \nonumber
z_1 = \frac{T_1}{T}, \;\; T_1= \frac{\mu_{imp} \mu_n \kappa}{2 r_0^3}.
\end{eqnarray}
Here $n_e$ is the electronic density $n_e = \frac{k_F^3}{3 \pi^2}$.

In analogy to (\ref{7}) one obtains the impurity contribution to the specific heat at zero external magnetic field
\begin{eqnarray}
\label{15}
C-C_0 = \frac{2\pi}{3} \frac { n_{imp} \mu_{imp} n_n \mu_n \kappa }{T} I_3(T), \\ \nonumber
I_3(T) = \int_0^\infty \frac{dz}{z^2} \frac{y^2}{\cosh^2 y}.
\end{eqnarray}

Here  the parameter $y = y(z)$ is defined in (\ref{13}).
Let us now give an analysis of the expressions for $\chi$ (\ref{12})
 and $C$ (\ref{15}) in different limiting cases.
 
In simple metals like $Pt$ the electron density $n_e$ (\ref{14}) is equal to the density
of nuclei and is of order of $a^{-3}$, where $a$ is the lattice spacing. 
For the clean case (no nonmagnetic impurities) the mean free path $r_0 \gg a$. Then 
from the formulas (\ref{14}) we find that 
there exists an extended
region of $T$, $T_1 \ll T \ll T_0$, such that one can assume 
$z_1=0$, $z_0 = \infty$. This corresponds to $I_2 = 0.723 $ and $I_3 = 0.636 $.
We see that in this region the  values of $I_2$  
and $I_3$  
do not depend on $T$, leading to $C \sim 1/T$ and $\chi \sim 1/T^2$. The oscillating behaviour of the RKKY
interaction does not show itself in this region, the oscillations of $V_{RKKY}$ are averaged out.
The graphs of $I_2(z_0)$, $I_3(z_0)$ for
$z_1 = 0$ are shown in the Fig. 1a, 1b. One can see from the Fig. 1a, 1b that the values  of $I_2(z_0)$, $I_3(z_0)$
oscillate only by about two times in the whole region $0 < z_0 < \infty$. The asymptotic value is approached
slowly, since  $I_2$ and  $I_3$ depend effectively on $z_0^{1/3}$. 
Note that for the nuclear specific heat of $PtFe_x$ 
the $C \sim 1/T$ law extends through  four orders of magnitude of $T$ \cite{wendler}.

For a dirty metal there exists a temperature region $T_{cn} \ll T \ll T_1 < T_0 $ (\ref{14}), where on one hand
the nuclear spins are uncorrelated, and on the other hand  one needs to account for the suppression of the
RKKY interaction (\ref{2})  related to the finite electron mean free path $r_0$. In this region of temperatures
one has $z_0 = \infty$ and  $z_1 \gg 1$ (\ref{13}) and the parameters $I_2$, $I_3$ (\ref{12},
\ref{15}) have the 
asymptotics 
\begin{equation}
\label{16}
I_2  \simeq 0.09 \frac{1}{z_1} \ln^3 z_1,  \; I_3 \simeq 0.25 \frac{1}{z_1} \ln^2 z_1.
\end{equation}

Therefore, for $T \ll T_1$, a saturation of the RKKY contribution to the impurity part of $C$ and $\chi$ 
occurs. 
\begin{equation}
\label{17}
C - C_0 \sim \frac{\kappa}{T_1} \ln^2 \frac{T_1}{T},	 \; \chi - \chi_0 \sim \frac{\kappa}{T T_1} \ln^3 \frac{T_1}{T}
\end{equation}

The graphs of $I_2, I_3$ as functions of $z_1$ for $z_0 = \infty$ are given in the Fig. 2a, 2b. 

Let us now go further down in temperatures. For a dirty metal, when the temperature is lowered, a situation occurs, where the saturated RKKY contribution (\ref{2},\ref{17}) to
$C$ and $\chi$, which is proportional to the large parameter $\kappa$, becomes lower, than the dipole-dipole
interaction (\ref{1},\ref{9}), which is proportional to $1/T$. This happens for $T \ll \frac{T_0}{\kappa}$. 
Increasing the concentration of nonmagnetic impurities one can narrow the temperature window $T_1 \ll T \ll T_0$, where the strong RKKY interaction plays role, and extract the contribution of  the magnetic dipole-dipole
interaction (\ref{1}). Therefore, taking into account that $\kappa \gg 1$, we await a very strong dependence of the impurity part of $C$ and $\chi$ on the concentration of the nonmagnetic impurities in the metal.

As a conclusion, there exist two distinct temperature regions, where the law $C - C_0 \sim 1/T$ is fulfilled.
 In the first region, $T_1 \ll T \ll T_0$, the impurity contribution to $C$ is related to the RKKY interaction.  
 In the second region, $T_{cn}
 \ll T \ll \frac{T_1}{\kappa}$, the main contribution to $C(T)$ is given by the magnetic dipole-dipole interaction  
(\ref{1})
. In the interval of temperatures $\frac{T_1}{\kappa} < T < T_1$  the
impurity contribution to $C - C_0$ has only a weak logarithmic temperature dependence. Measuring 
the impurity contribution to $C$ and $\chi$ one can  extract the parameter
$\kappa$ of the RKKY interaction (\ref{2}).

{\bf Acknowledgements}

Two of us (A. M. D. and Yu. N. O.) wish to acknowledge die Max-Planck-Gesellschaft 
zur F\"orderung der Wissenschaften and Max-Planck-Institut f\"ur Physik komplexer
Systeme for the hospitality during the period of doing  this work.

The research of A. M. Dyugaev was supported by the INTAS-RFBR 95-553 grant.

The research  of Yu. N. Ovchinnikov was supported by the CRDF grant RP1-194 and 
by the Naval Research Lab contract N 00173-97-P-3488.

\newpage

\noindent
FIGURE CAPTIONS
\\
\\
\\
Fig. 1a:
$I_2$  as a function of  $z_0$ for $z_1 = 0$.
\\
\\
Fig. 1b:
$I_3$  as a function of  $z_0$ for $z_1 = 0$.
\\
\\
Fig. 2a:
$I_2$  as a function of  $z_1$ for $z_0 = \infty$.
\\
\\
Fig. 2b:
$I_3$  as a function of  $z_1$ for $z_0 = \infty$.

\end{document}